\documentclass[twocolumn,superscriptaddress,preprintnumbers,amsmath,amssymb,floatfix]{revtex4-1}
\pdfoutput=1
\usepackage{graphicx} 
\usepackage{dcolumn} 
\usepackage{bm}
\usepackage{color}
\usepackage{hyperref}
\usepackage{graphicx}
\usepackage{amsmath}
\usepackage{setspace}

\begin{document}
\title{
Impact of quadrupole deformation on intermediate-energy heavy-ion collisions}
\author{Xiao-Hua Fan}
\affiliation{School of Physical Science and Technology, Southwest University, Chongqing 400715, China}
\affiliation{RIKEN Nishina Center, Wako, Saitama 351-0198, Japan}

\author{Zu-Xing Yang}
\email{zuxing.yang@riken.jp}
\affiliation{RIKEN Nishina Center, Wako, Saitama 351-0198, Japan}
\affiliation{School of Physical Science and Technology, Southwest University, Chongqing 400715, China}

\author{Peng-Hui Chen}
\affiliation{College of Physics Science and Technology, Yangzhou University, Yangzhou, Jiangsu 225002, China}

\author{Shunji Nishimura}
\affiliation{RIKEN Nishina Center, Wako, Saitama 351-0198, Japan}

\author{Zhi-Pan Li}
\affiliation{School of Physical Science and Technology, Southwest University, Chongqing 400715, China}

\begin{abstract}

This study employs the isospin-dependent Boltzmann-Uehling-Uhlenbeck model to simulate intermediate-energy heavy-ion collisions between prolate nuclei $^{24}$Mg.  
The emphasis is on investigating the influence of centrality and orientation in several collision scenarios. 
The final-state particle multiplicities and anisotropic flows are primarily determined by the eccentricity and the area of the initial overlap.
This not only provides feedback on the collision systems, but also, to some extent, provides a means to explore the fine structure inside deformed nuclei.
Additionally, non-polarized collisions have been further discussed.
These results contribute to the understanding of the geometric effects in nuclear reactions, and aid in the exploration of other information on reaction systems, such as the equation of state and nuclear high-momentum tail.

\end{abstract}

\maketitle

\section{Introduction}

Most open-shell nuclei are deformed in their ground state, which originates from collective motion induced by the interaction among valence nucleons \cite{Moeller2016At.DataNucl.DataTables109110.1204, Heyde2011Rev.Mod.Phys.83.1467}.
The multipole deformation researches have been conducted widely by various phenomenological and microscopic models \cite{Troltenier1991.105128, Moeller2016At.DataNucl.DataTables109110.1204, Reinhard2022Phys.Rev.C106.014303, Reinhard2021Comput.Phys.Commun.258.107603, Stoitsov2005Comput.Phys.Commun.167.4363, Zhang2022At.DataNucl.DataTables144.101488, Pan2022Phys.Rev.C106.014316, Niksic2008Phys.Rev.C78.034318}. 
Recently, deep neural network-based generative models are also being developed for producing potential energy surfaces, rotational inertia, and vibrational inertia of deformed nuclei, which can be further utilized to investigate excitation spectra within the five-dimensional collective Hamiltonian approach \cite{Lasseri2020Phys.Rev.Lett.124.162502, Akkoyun2013PhysicsofParticlesandNucleiLetters10.528534}.
Experimentally, quadrupole deformation can be measured from the rotational spectra of the nuclear excited state or the electric quadrupole moments derived from the hyperfine splitting of the atomic spectral line \cite{Moeller2016At.DataNucl.DataTables109110.1204}.

Heavy-ion collisions are another effective means of investigating nuclear properties.
At various energies, it is proven that emitted particles and gamma rays carry information about the properties of the individual nuclei and the composite systems \cite{Gao2023Phys.Lett.B838.137685, Yue2022Universe8.491}, such as equation of state (EoS) \cite{Tsang2009Phys.Rev.Lett.102.122701, Jhang2021Phys.Lett.B813.136016, Fan2018Phys.Rev.C97.034604}, short-range correlations between nucleons \cite{Yang2018Phys.Rev.C98.014623, Cai2022Phys.Rev.C105.064607}, medium effects in scattering cross sections \cite{Yong2017Phys.Rev.C96.044605, Li2022Phys.Lett.B828.137019}, and nuclear bubble or hollow configurations \cite{Fan2019Phys.Rev.C99.041601}.
A series of critical researches have been carried out via ultra-relativistic quantum molecular dynamics model \cite{Deng2022Phys.Lett.B835.137560, Saito2021Eur.Phys.J.A57., Kundu2021Phys.Rev.C104.024907}, neural network model \cite{Li2021Phys.Rev.C104.034608,Wang2021Phys.Lett.B822.136669},  antisymmetrized molecular dynamics model \cite{Frosin2023Phys.Rev.C107.044614, Takatsu2023Phys.Rev.C107.024314}, the heavy-ion phase space exploration model \cite{Frosin2023Phys.Rev.C107.044614}, and dinuclear system model \cite{Niu2021Nucl.Sci.Tech.32.}.
On the other hand, relativistic heavy-ion collisions offer an opportunity to study nuclear deformation since the collision dynamics are dominated by the interactions between the individual nucleons at high incident energies, rather than the collective motion of the entire nucleus or inter-nucleon correlations \cite{Adamczyk2015Phys.Rev.Lett.115.222301, Sirunyan2019Phys.Rev.C100.044902, Acharya2018PhysicsLettersB784.8295, Aad2020Phys.Rev.C101.024906}. 
As early as 2000, a pioneering work for deformed heavy-ion collisions was conducted by Bao-An Li \cite{Li2000Phys.Rev.C61.021903}.
In recent years, the final-state anisotropic flows are sensitive to the strengths of multipole components of the nucleon distribution in transverse plane was found via a parton-cascade based multiphase transport model (AMPT)\cite{Zhang2005Phys.Rev.C72.024906, Giacalone2021Phys.Rev.Lett.127.242301, Zhang2022Phys.Rev.Lett.128.022301, Jia2022Phys.Rev.C105.014905, Jia2022Phys.Rev.C105.014906}.
However, the high cost of high-energy reactions means that the simulation and study of low- to medium-energy reactions remain necessary.
Moreover, it is worth exploring how the fine structure inside the deformed nucleus would affect the final-state.
More importantly, exploring geometric effects can also enhance the understanding of other properties in reaction systems.

In this study, we will simulate intermediate-energy heavy-ion collisions between deformed $^{24}\text{Mg}$ nuclei in several collision scenarios by using the widely accepted isospin-dependent Boltzmann-Uehling-Uhlenbeck (IBUU) transport model.
The structure of this article will be organized as follows.
In Sec.~\ref{sec2}, we will introduce the initializations for IBUU model, including the adjustment on impact parameters and collision orientations.
We will further analyze the variation of the central density, the yield of emitted particles, and the collective flows to speculate on the mechanism of the ultracentral, semi-central, and peripheral polarized collision of deformed nuclei in Sec.~\ref{sec3}.
Subsequently, we will take into account the rotations of the projectile in Sec.~\ref{sec:4}, which will provide guidance for event-by-event analysis of the orientation of the projectile and the target in experiments.
In Sec.~\ref{sec:5}, the non-polarized collisions will be further investigated.
Finally, we will summarize briefly in Sec.~\ref{sec:6}

\section{Initialization of IBUU transport model \label{sec2}}

The Monte Carlo method-based IBUU model simulates the physical processes involved in the phase-space evolution of baryons and mesons during heavy-ion collisions, including elastic and inelastic scattering, absorption, and decay of particles \cite{Bertsch1988Phys.Rep.160.189233}.
The version we employed \cite{Yang2021J.Phys.GNucl.Part.Phys.48.105105, Yong2016Phys.Rev.C93.014602, Yang2018Phys.Rev.C98.014623,Guo2019Phys.Rev.C100.014617,Cheng2016Phys.Rev.C94.064621} has incorporated the Coulomb effect, Pauli blocking, medium effects on scattering cross sections, etc.  
To minimize the fluctuations, this work simulates and averages 300,000 events for each collision scenario.

Typically, the deformation geometric effects on the final-state are studied by initializing the nucleon spatial distribution, with the traditional deformed Woods-Saxon density being widely adopted  \cite{Jia2022Phys.Rev.C105.014905, Zhang2022Phys.Rev.Lett.128.022301}.
In this study, replacing the Woods-Saxon form, we select a relativistic mean field  (RMF)-based self-consistent calculation with the point-coupling DD-PC1 interaction  \cite{Niksic2008Phys.Rev.C78.034318} including the Bardeen-Cooper-Schrieffer pairing to initialize $^{24}\text{Mg}$.
The RMF calculations have gained wide acceptance due to the capability in accurately reproducing the masses and root-mean-square radii of deformed nuclei \cite{Niksic2008Phys.Rev.C78.034318}.
This indicates that the corresponding density distribution captures nuclear structural information in detail, enabling more realistic collision simulations to be performed.
As a light, stable, and highly deformed nucleus \cite{Moeller2016At.DataNucl.DataTables109110.1204, Zhang2022At.DataNucl.DataTables144.101488, Pan2022Phys.Rev.C106.014316, Reinhard2021Comput.Phys.Commun.258.107603}, $^{24}\text{Mg}$ is a favorable choice for heavy-ion collision experiments to study the deformation effect. 
Meanwhile, $^{24}\text{Mg}$, as a symmetric nucleus, can mitigate the instability caused by the symmetry energy.

\begin{figure}[tb]
\includegraphics[width=8.5 cm]{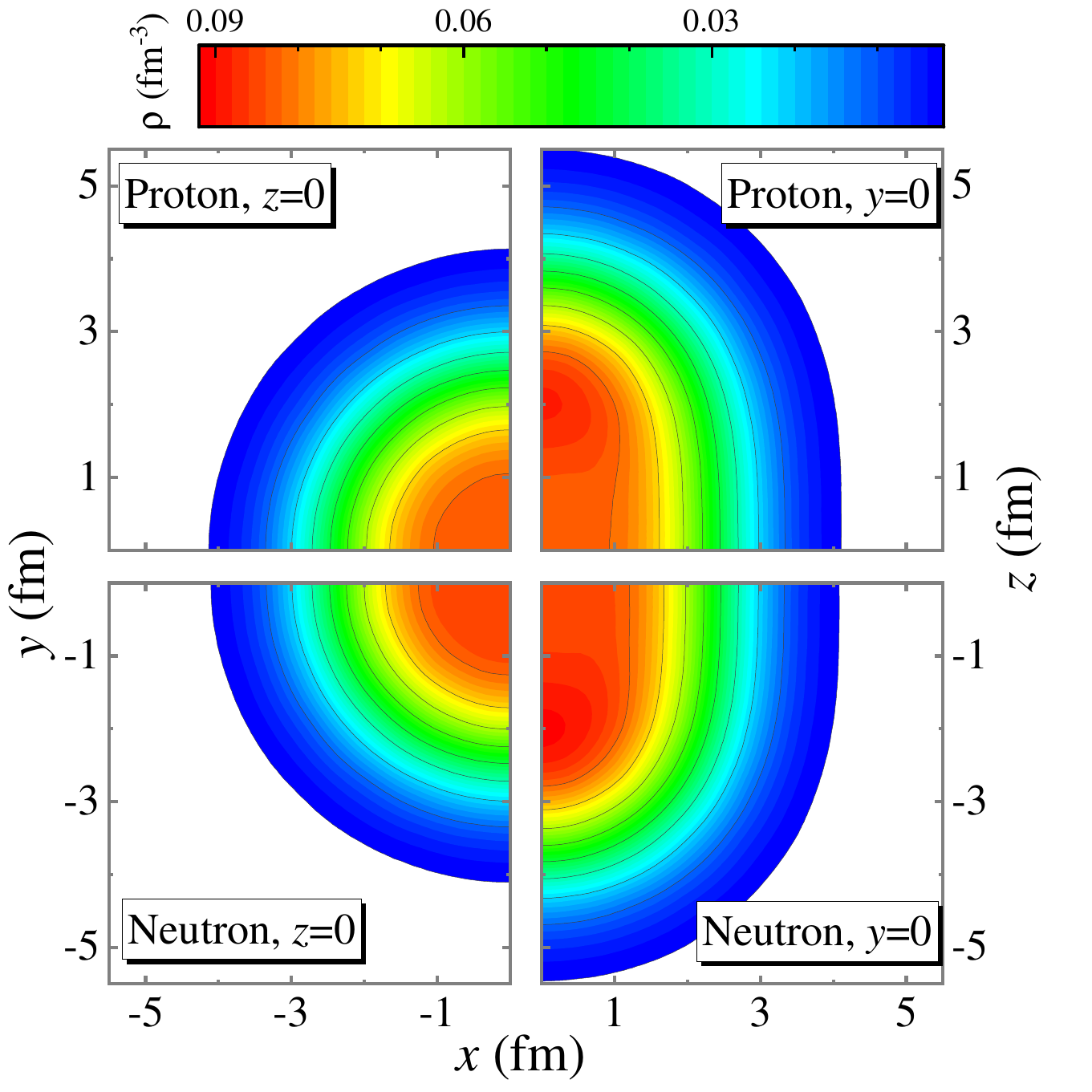}
\caption{\label{fig1} The neutron and proton densities for $^{24}\text{Mg}$ in the ground state in the $x-y$ ($z=0$) and $x-z$ ($y=0$) planes calculated by relativistic mean field with DD-PC1 interaction.
  }
\end{figure}

Figure~\ref{fig1} shows the neutron and proton densities for $^{24}\text{Mg}$ in the transverse $x-y$ section and longitudinal $x-z$ section through the center, with variations in color representing the densities. 
The density distributions of both protons and neutrons in the $x-y$ plane exhibit radial symmetry, whereas those in the $x-z$ plane are elliptical, indicating that $^{24}\text{Mg}$ is a prolate ellipsoidal nucleus.
Based on the mass quadrupole moments, quadrupole deformation $\beta_2$ is defined in detail as \cite{Horiuchi2012Phys.Rev.C86.024614}:
\begin{equation}
\beta_2 = \sqrt{\beta_{2,0}^2 + \beta_{2,2}^2},
\end{equation}
where
\begin{equation*}
\begin{aligned}
\beta_{2,0} &= \sqrt{\frac{\pi}{5}} \frac{\langle 2z^2-x^2-y^2 \rangle}{\langle r^2 \rangle},\\
\beta_{2,2} &= \sqrt{\frac{3\pi}{5}} \frac{\langle y^2-x^2 \rangle}{\langle r^2 \rangle}.
\end{aligned}
\end{equation*}
Calculations show that triaxial deformation $\beta_{2,2}=0$, and thus $\beta_2 =\beta_{2,0} = 0.53$.
Another point worth noting is that there are two peaks ($z = \pm 2\, \text{fm}$) in the density variation along the $z-$direction, as shown in the right panels.
Under the current densities, the local density approximate Fermi gas method is employed to generate the momentum distributions.

\begin{figure}[tb]
\includegraphics[width=8 cm]{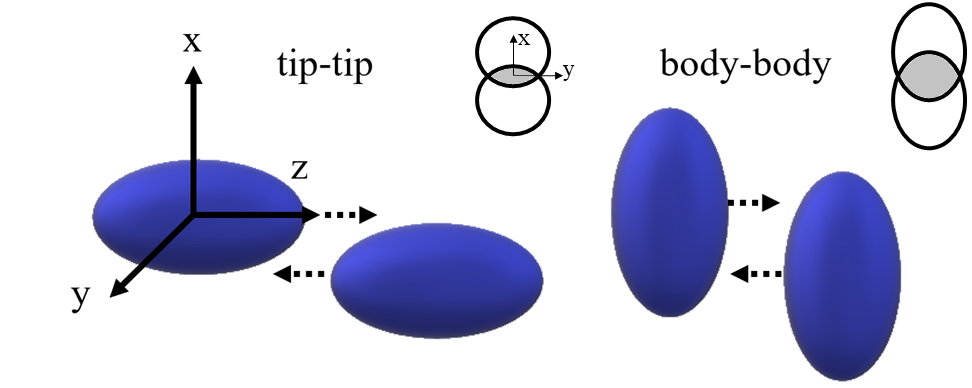}
\caption{\label{fig2} Simulated schematic for tip-tip collision and body-body collision of $^{24}\text{Mg}$ + $^{24}\text{Mg}$, where $x-z$ plane is the reaction plane. 
In the upper right corner, resulting transverse areas of overlap are shown.
  }
\end{figure}

Within the framework of the IBUU transport model, the two extreme polarization states of the reacting nuclei are comprehensively discussed through ultracentral, semi-central, and peripheral collisions of $^{24}\text{Mg}$ + $^{24}\text{Mg}$ at 0.5 GeV/nucleon. 
The collision schemes with the two polarizations are illustrated in Fig.~\ref{fig2}, labeled as tip-tip collision (left panel) and body-body collision (right panel), respectively.
In the upper right corner, resulting transverse areas of initial overlap are shown, where the $x-$direction represents the direction of the impact parameter $b$.
The ultracentral, semi-central, and peripheral collisions correspond to $b=0\,\text{fm}$, $b=4\,\text{fm}$, and $b=6\,\text{fm}$, respectively.
In fixed target experiments, the difference in collision centrality can be determined via spectators from the projectiles.
With the assistance of machine learning methods, it is possible to obtain more accurate information about impact parameters \cite{Li2020J.Phys.GNucl.Part.Phys.47.115104}.
These schemes allow, to some extent, for the generalization and discussion of reactions of deformed nuclei under non-polarized conditions.
As a comparison, nucleon density distributions constrainingly calculated at $\beta_2 = 0$ point are also selected to initialize the spatial distribution, which is labeled as spherical-spherical (sph.-sph.) collision.
Furthermore, the changes in final-state anisotropic flows and emitted particle yields caused by the orientation of the projectile nucleus are also extensively discussed in Sec.~\ref{sec:4}.

\section{Analysis and discussion \label{sec3}}

\begin{figure}[tb]
\includegraphics[width=9 cm]{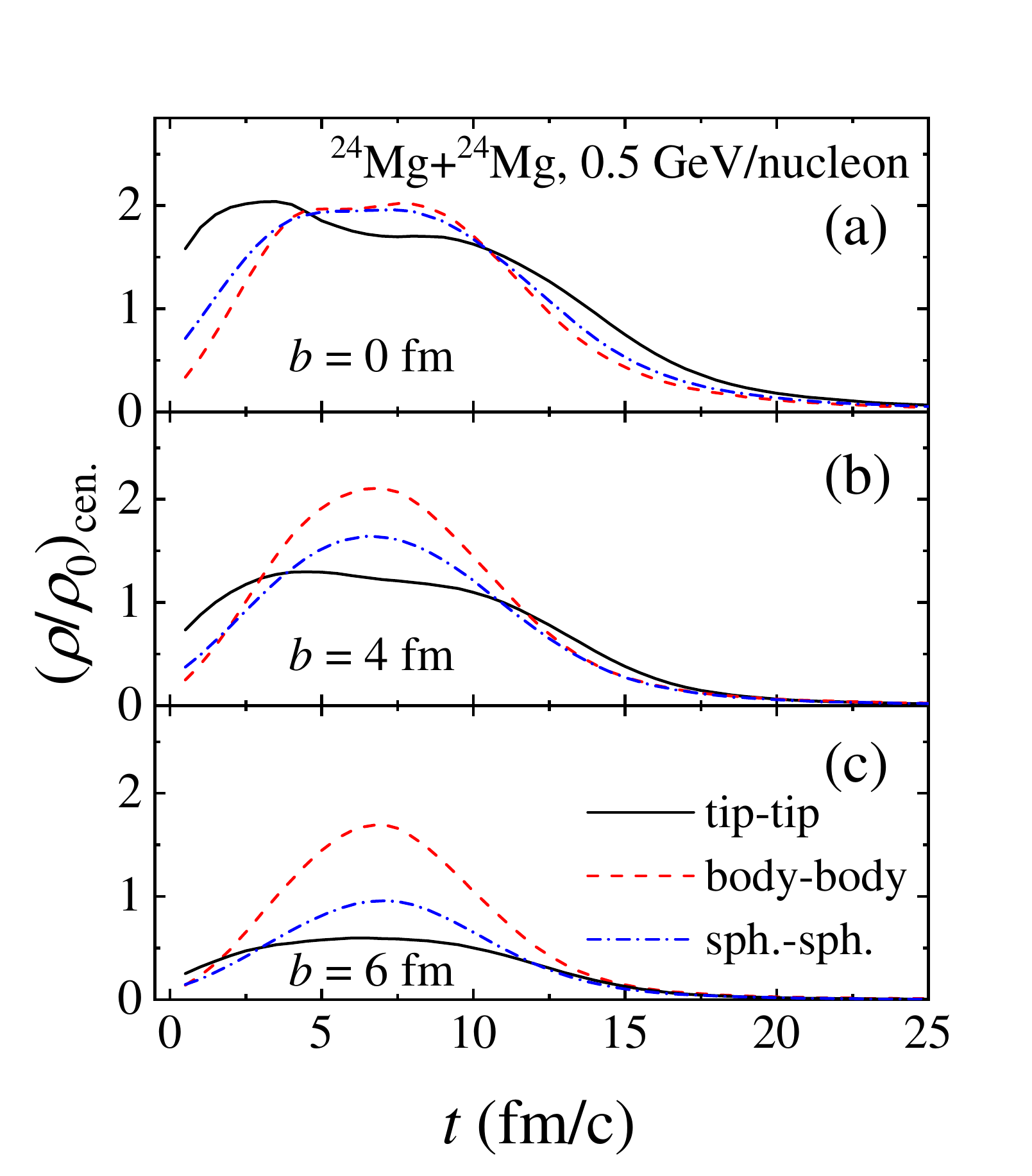}
\caption{\label{fig3} The central baryon density as a function of time for the ultracentral, semi-central, and peripheral collisions of $^{24}\text{Mg}$ + $^{24}\text{Mg}$ at 0.5 GeV/nucleon with three collision scenarios.
}
\end{figure}

As the nuclear pressure plays a crucial role in determining the reaction products, the exploration of the highest achieved central baryon density is prioritized, as illustrated in Fig.~\ref{fig3}.
The results for ultracentral collisions are presented in panel (a).
It can be observed that in all three collision scenarios, the central densities reach twice the saturation density, where the saturation density is $\rho_0 = 0.17\,\mathrm{fm}^{-3}$.
When comparing body-body collisions to sph.-sph. collisions, no apparent variations are found, as both types of collisions reach the same highest density and maintain similar durations.
For tip-tip collisions, the baryons are squeezed earlier and undergo two drops in density, which interestingly corresponds to the existence of two peaks along the $z-$axis (see Fig.~\ref{fig1}). 
Such structure can be more comprehensively studied through proton-induced reactions with the aid of target nucleus polarization, as shown in previous research on the $^{48}$Si bubble configurations \cite{Fan2019Phys.Rev.C99.041601}. 
Similarly to $^{48}$Si, $^{24}$Mg can also be interpreted as a bubble configuration present in the central region along the $z-$axis.
In Sec.~\ref{sec:4}, we will also discuss the impact of this bubble on observables from a different perspective.

The scenarios with increasing impact parameters are shown in (b) semi-central collisions and (c) peripheral collisions.
Notably, body-body collisions with larger impact parameters almost reach the same high density (pressure) as ultracentral collisions.
This differs from sph.-sph. and tip-tip collisions with obviously lower densities, where the peripheral tip-tip collisions only achieve half of the saturation density.
The centrality-induced pressure differences can be attributed to the initial density distribution, which ultimately leads to significant influences in the final-state observables.

\begin{figure}[tb]
\includegraphics[width=9 cm]{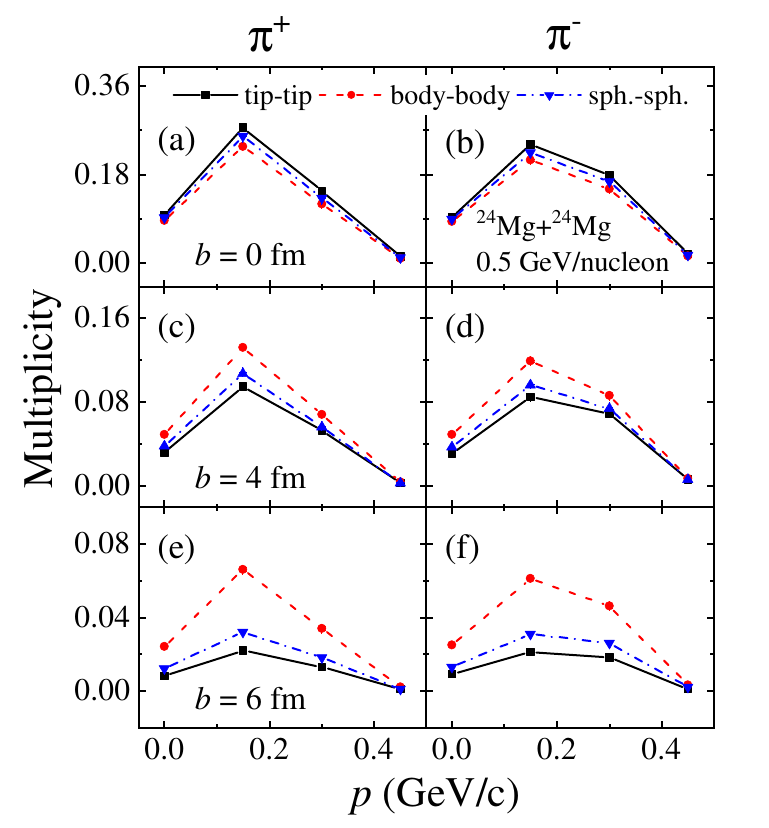}
\caption{\label{fig4} Multiplicities of $\pi^-$ and $\pi^+$ mesons as a function of momentum for the ultracentral, semi-central, and peripheral collisions of $^{24}\text{Mg}$ + $^{24}\text{Mg}$ at 0.5 GeV/nucleon with three collision scenarios.
  }
\end{figure}

Figure~\ref{fig4} displays the momentum-dependent multiplicities of $\pi^-$ and $\pi^+$ mesons, which serve as reliable probes of the central density.
For ultracentral collisions, as depicted in panels (a) and (b), the three collision scenarios are not clearly distinguished. 
This can be explained by the fact that in these collisions, all nucleons participate in the reaction with similar nuclear pressure and durations.
Increasing the impact parameters leads to an increase in spectators and a decrease in nuclear pressure, causing an overall reduction in the multiplicities as shown in panels (c)-(f), especially for sph.-sph. and tip-tip collisions. 
Thus, the increased $b$ amplifies the disparity among the three collision scenarios.
The centrality is experimentally defined as the overlap of the two colliding nuclei as a percentage of the total cross-section.
In this sense, the body-body case obviously has the greatest centrality with the same $b$ as other cases.
Experimental investigations should focus on the momentum of 0.15 GeV/c since it consistently yields the highest multiplicities across all impact parameters and orientations, and the dissimilarities among collision scenarios are most pronounced at this momentum.
It is also noteworthy that the multiplicities of neutrons and protons are studied and exhibit similar trends and patterns.

\begin{figure}[tb]
\includegraphics[width=9 cm]{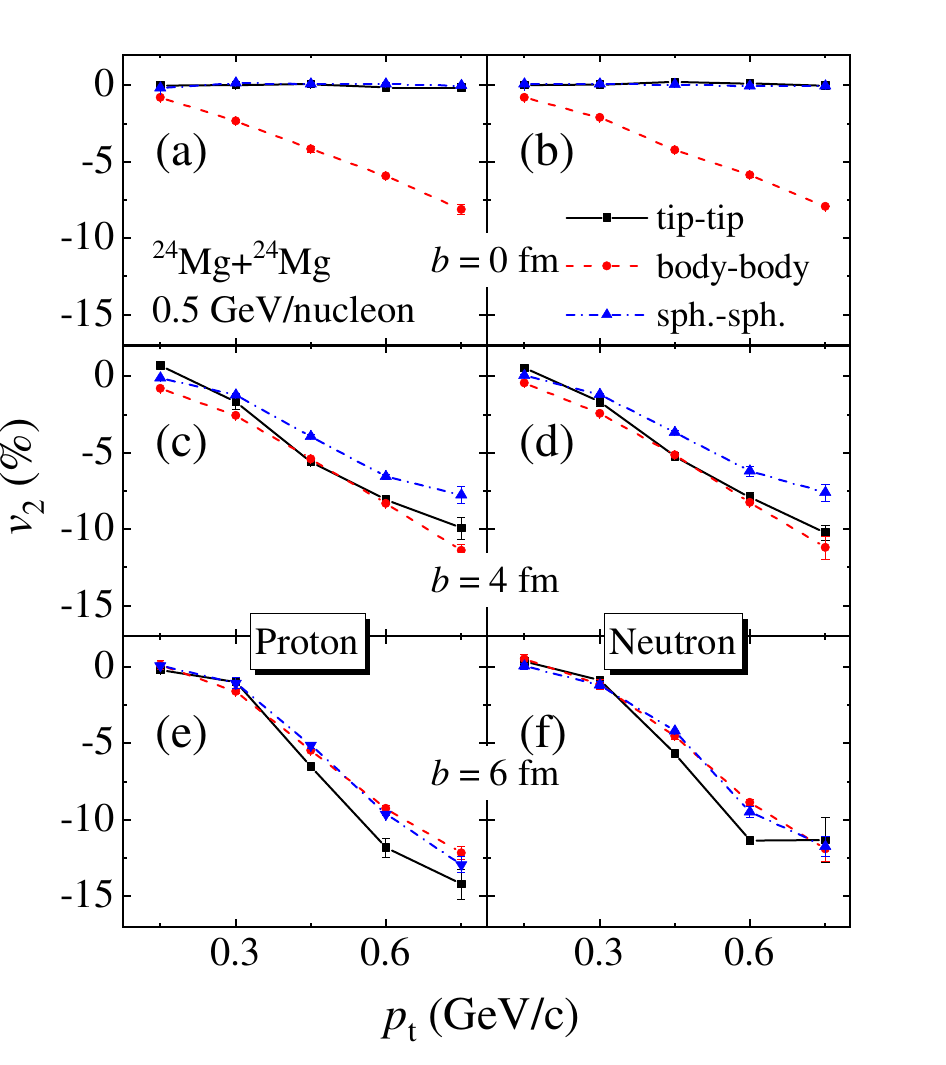}
\caption{\label{fig5} The proton and neutron elliptic flows as a function of transverse momentum 
 $p_t(= \sqrt{p^2_x+p^2_y})$ at $(y/y_\text{beam})_\text{c.m.} \leq 0.3$ for the reaction of $^{24}\text{Mg}$ + $^{24}\text{Mg}$ at a beam energy of 0.5 GeV/nucleon with three collision scenarios.
  }
\end{figure}

To cancel out nuclear pressure from dense matter to some extent, we now focus on studying nucleon elliptic flow, which has been demonstrated as a sensitive probe of the quadrupole deformation of the initial state in relativistic heavy-ion collisions \cite{Zhang2022Phys.Rev.Lett.128.022301, Giacalone2021Phys.Rev.Lett.127.242301}.
Elliptic flow is characterized by the second-order anisotropic flow coefficient $v_2$, which can be expressed as:
\begin{equation}
v_2 = \left\langle \mathrm{cos}(2\phi) \right\rangle,
\end{equation}
where $\phi = \arctan{(p_y/p_x)}$ represents the transverse azimuth angle.
As a setting, we uniformly intercept the collective flows with the central rapidity of $(y/y_\text{beam})_\text{c.m.} \leq 0.3$.
The final-state elliptic flows for the three collision scenarios are shown in Fig.~\ref{fig5}.
In ultracentral collisions (a)-(b), the corresponding elliptic flows are naturally absent due to the lack of anisotropy in the transverse plane for both tip-tip and sph.-sph. collisions, which results in an appreciable divergence from body-body collisions.
Such divergence is particularly evident in the energetic emitted nucleons in the transverse direction.
As the impact parameters gradually increase (c)-(f), the anisotropy in the transverse plane becomes non-negligible for tip-tip and sph.-sph. collisions manifested as a substantial enhancement in elliptic flows.
In detail, we can always notice that tip-tip collisions have stronger elliptic flow than sph.-sph. collisions.
This is caused by the initial eccentricities of the overlap, determined by \cite{Zhang2022Phys.Rev.Lett.128.022301, Jia2022Phys.Rev.C105.014905, Alver2007Phys.Rev.Lett.98.242302, Sorge1999Phys.Rev.Lett.82.20482051}
\begin{equation}
\label{eq_3}
\epsilon_n = \mid\frac{\int r_\perp^n e^{in\phi_\perp}\rho(r_\perp,\phi_\perp) r_\perp dr_\perp d\phi_\perp }{\int r_\perp^n \rho(r_\perp,\phi_\perp) r_\perp dr_\perp d\phi_\perp} \mid
\end{equation}
where $\rho(r_\perp,\phi_\perp)$ is the transverse density in the overlap.
According to the positive correlation ($v_2 \varpropto \epsilon_2$) \cite{Jia2022Phys.Rev.C105.014905}, peripheral tip-tip collisions with a larger initial eccentricity in the overlap region naturally produce stronger elliptic flows, compared with sph.-sph. collisions.
However, an anomaly can be observed in the figures: as $b$ increases, the initial eccentricity of the overlap theoretically decreases for body-body collisions, but the strength of the elliptic flow remains unchanged.
This distinction physically separates intermediate-energy reactions from high-energy reactions. 
In high-energy reactions, the reaction time is shorter and the influence of spectators is limited, so geometric effects can be more cleanly captured. 
However, in intermediate-energy reactions, the longer reaction with the viscosity among nucleons also has a significant impact on the results.
At this point, the dilute nucleons dispersed in the periphery of the nucleus also play a great role in $v_2$.
This may imply the breakdown of the linear dependence of $v_2$ on $\epsilon_2$, instead being replaced by even stronger anisotropic flow.

\section{The orientation effect of the projectile \label{sec:4}}

\begin{figure}[tb]
\includegraphics[width=9 cm]{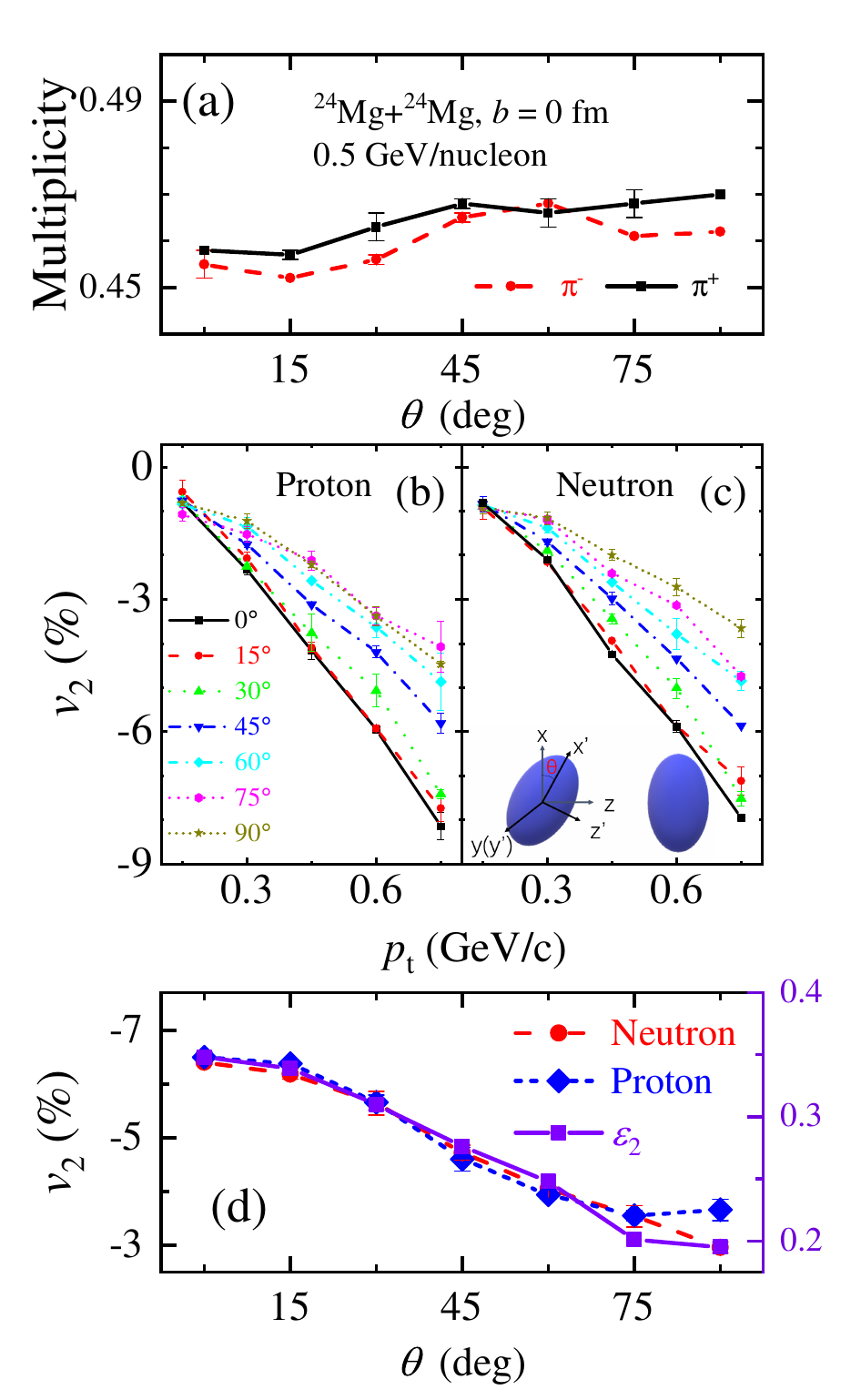}
\caption{\label{fig6} The orientation effects of the projectile for the ultracentral collisions of $^{24}$Mg + $^{24}$Mg at 0.5 GeV/nucleon. 
Panel (a): Multiplicities of $\pi^-$ and $\pi^+$ mesons as a function
of the orientation of the projectile.
Panels (b) and (c): The proton and neutron elliptic flows as a function of transverse momentum with different orientations of the projectile.
The schematic diagram of the orientation is shown in the lower left corner of panel (c).
Panel (d): $v_2$ ($p_t \geqslant 0.5~\text{GeV/c}$) and $\epsilon_2$ as functions of the orientation $\theta$.
  }
\end{figure}

In this section, we consider the scenario where ultracentral collisions occur between a polarized target and an unpolarized projectile.
In such a scenario, two situations may arise: 1.~the projectile fully hits the target, where all nucleons from the projectile are participants; 2.~the projectile partially hits the target, where some nucleons from the projectile as spectators fly off after the collision with essentially unchanged velocity.
In experiments, the spectators can be distinguished on an event-by-event basis \cite{Bertsch1988Phys.Rep.160.189233}, allowing us to concentrate on the former.
The scenarios where the projectile fully hits the target are shown in Fig.~\ref{fig6}, with the lower left corner of panel (c) displaying the schematic diagram of the orientation.
Starting from the body-body case, a rotation operator around the $y-$axis in terms of Euler angles ($\Omega(0,\theta,0)$) is applied to the intrinsic coordinate system $xyz$ of the projectile. 
The rotated coordinate system is denoted as $x'y'z'$. 
In the special case where $\theta=90^\circ$, the system evolves into the tip-body scenario.

The multiplicities of $\pi^-$ and $\pi^+$ mesons as a function of the orientation of the projectile are exhibited on panel (a).
Clearly, the corresponding yield of mesons shows modest dependence on $\theta$, even though the participants from the target decrease as $\theta$ increases in principle.
In fact, the multiplicities of free protons and neutrons have also been examined and found the same low sensitivity to $\theta$.
This means that these observables remain good probes for testing other properties of nuclear systems, such as the symmetry energies and high-momentum tails.

The effects of orientation on proton and neutron elliptic flows are displayed in panels (b) and (c), respectively.
It is evident that the elliptic flows weaken with increasing verticality, which can be geometrically explained by the initial eccentricity of the overlap.
In detail, more noticeable changes are observed at higher momenta for both protons and neutrons, which can better assist us in inferring the orientation of the projectile based on experiments.
The dashed line and solid line shown in panel (d) represent the changes of $v_2$ ($p_t \geqslant 0.5~\text{GeV/c}$) and $\epsilon_2$ with respect to the orientation, respectively.
An increase in $\theta$ leads to a decrease in the corresponding eccentricity, then the strength of the $v_2$ weakens synchronously.
Meanwhile, the most noticeable sensitivity occurs around $45^\circ$, whereas the sensitivity is relatively lower near $0^\circ$ and $90^\circ$, which reflects that the gradient of eccentricity variation is greatest around $45^\circ$.
It must also be emphasized that the $v_2$ and the $\epsilon_2$ in panel (d) almost coincide, indicating that the linear response relation in the ultracentral collisions at intermediate energies is still reliable.
This can be further expressed as $|v_2| = k \epsilon_2$, where the response coefficient $k \approx 0.0675$ can be obtained from the special case of body-body collisions.
About the response coefficient more properties, including experimental measurements and environmental dependencies, have been discussed and understood in detail \cite{Bernhard2016Phys.Rev.C94.024907, Bernhard2019Nat.Phys.15.11131117, Nijs2021Phys.Rev.Lett.126.202301}.

Additionally, we should pay attention to the situation at $\theta=90^\circ$, where the geometric overlap eccentricity equals 0.195. 
However, as we set the central region to a constant density, i.e. mathematically smoothed out the peak at $z = \pm 2\, \text{fm}$, we observed a decline in eccentricity to 0.175.
According to the linear response relation, this also means that the strength of $v_2$ will simultaneously decrease by $10\%$. 
This provides a novel method to probe the fine structure inside deformed nuclei.

\begin{figure}[tb]
\includegraphics[width=9 cm]{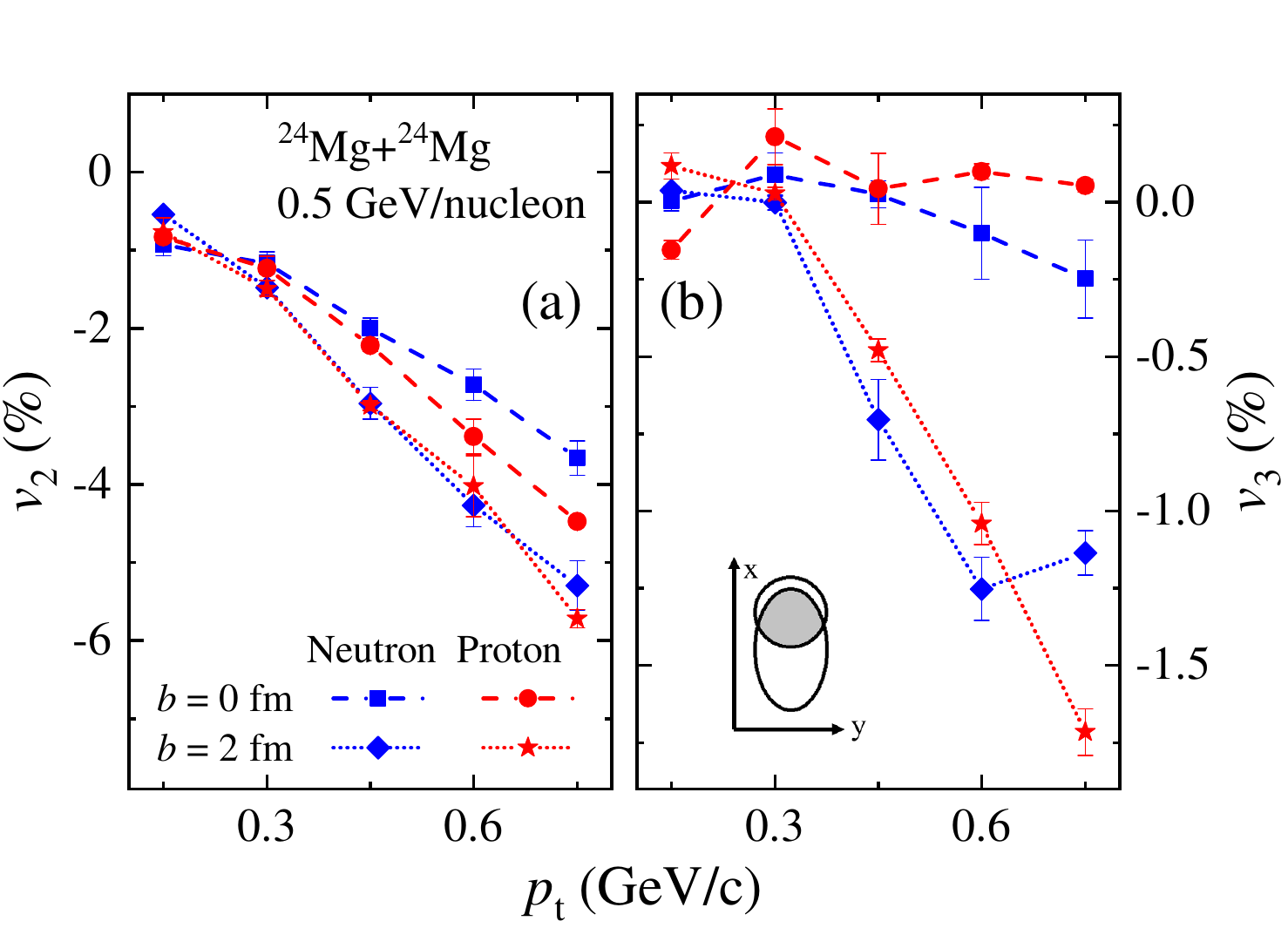}
\caption{\label{fig7} The proton and neutron elliptic (a) and triangular (b) flows as a function of transverse momentum 
 $p_t$ at $(y/y_\text{beam})_\text{c.m.} \leq 0.3$ for the reaction of $^{24}\text{Mg}$ + $^{24}\text{Mg}$ at a beam energy of 0.5 GeV/nucleon. 
In the lower-left corner, resulting transverse areas of overlap are shown.
  }
\end{figure}

Finally, within the tip-body case ($\theta = 90^\circ$), we further explore the influence of the impact parameters on collective flows.
The proton and neutron elliptic and triangular flows as a function of $p_t$ are shown in Fig.~\ref{fig7}.
As seen in panel (a), the elliptical flow strengthens slightly with increasing impact parameters, which is again due to the increased $\epsilon_2$.
A more discriminatory signal appears in (b) triangular flow $v_3$, especially at high $p_t$, which is defined as
\begin{equation}
    v_3 = \langle \cos(3\phi) \rangle.
\end{equation}
This can be analogously explained by $\epsilon_3$.
The initial cross-section diagram relating to $\epsilon_3$ is shown in the lower left corner of (b), where the triangular symmetry emerges with decreased centrality as shown by the shaded region.
It should be emphasized that, regardless of fluctuations, the collective flow induced by such symmetry will not be observed for a spherical nucleus in any case. 
Even for a quadrupole-deformed nucleus, triangular symmetry will not appear under ultracentral collisions.
Therefore, we can assert that $v_3$ at high $p_t$ is indeed a reliable and sensitive observable for deformation and centrality.

\section{Non-polarized collisions \label{sec:5}}

In the preceding discussions, we have examined the effects of deformed nuclei in various polarization scenarios. 
However, it is noteworthy that even in non-polarized conditions, the deformation effect can still be observed in ultra-relativistic heavy-ion collisions.
Therefore, further exploration of the deformation effect in non-polarized collisions within the intermediate energy regime is crucial.
To achieve this, the Euler rotation operator ($\Omega(\phi,\theta,0) = R_z(\phi)R_y(\theta)R_x(0)$) is applied independently to the target and the projectile, where only four degrees of freedom ($\theta_1,\phi_1,\theta_2,\phi_2$) are required for each ultracentral collision event due to the absence of triaxial deformation.
We average the results of these collisions with random orientation and denote it as Avg.
On the other hand, the spectators from the projectiles can be detected on an event-by-event basis in target shooting experiments.
In this manner, events lacking spectators from the projectiles can be readily screened out and averaged, labeled as Avg.2.
In these events, the projectile orientation ($\theta_2$, $\phi_2$) satisfy 
\begin{equation}
    \theta_2 \in 
\begin{cases}
    [\theta_1, 180^\circ-\theta_1] &  \theta_1 < 90^\circ\\
    [180^\circ-\theta_1, \theta_1] &  \theta_1 \geq 90^\circ
\end{cases}
\end{equation}
and 
\begin{equation}
    \phi_2 = \phi_1
\end{equation}
with $\theta_1 \in [0^\circ, 180^\circ]$ and $\phi_1 \in [0^\circ, 180^\circ]$ being the target orientation.
With continued technological advancements, the detection of spectators from the target will also be possible. 
This will provide additional assistance in determining the impact parameter from the narrow centrality bin of the current scenario. 
We hence set the impact parameter $b=0$ fm.

\begin{figure}[tb]
\includegraphics[width=9 cm]{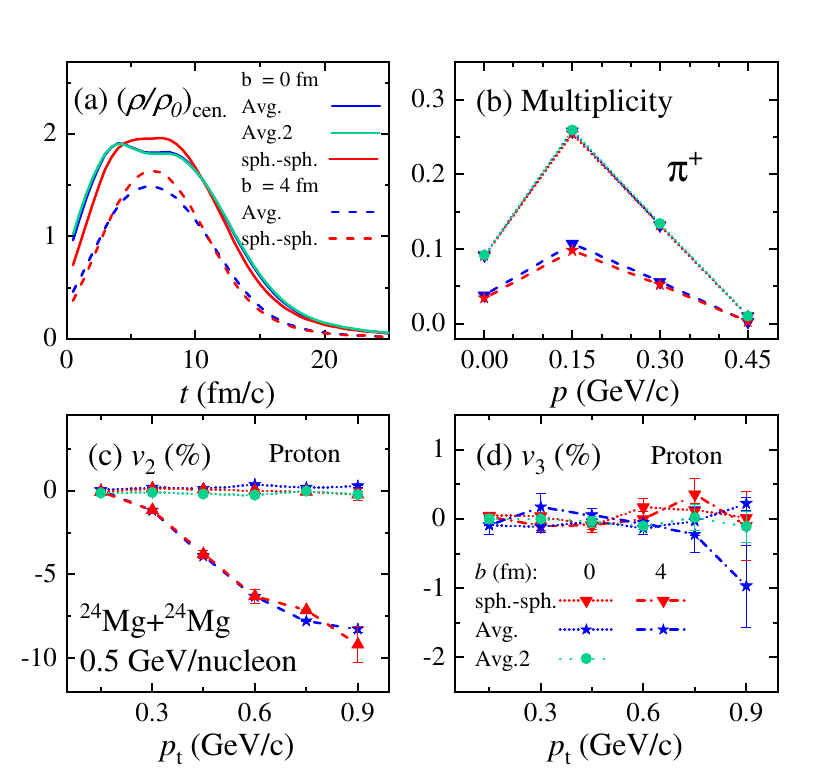}
\caption{\label{fig8} Comparisons between the average results--Avg. and Avg.2 (see text for details) of non-polarized collisions and sph.-sph. case for the reaction of ultracentral and semi-central $^{24}\text{Mg}$ + $^{24}\text{Mg}$ collisions at a beam energy of 0.5 GeV/nucleon. 
Panel (a): The central baryon density as a function of time. 
Panel (b): Multiplicities of $\pi^+$ meson as a function of momentum.
Panel (c): The proton elliptic flows as a function of
transverse momentum.
Panel (d): The proton triangular flows as a function of
transverse momentum.
The legend shared in panel (d) is applicable to panels (b)-(d).
  }
\end{figure}

Figure~\ref{fig8} showcases comparisons among the results of non-polarized collisions (Avg. and Avg.2) and the sph.-sph. scenario, including central density $(\rho/\rho_0)_\text{cen.}$, $\pi^+$ multiplicity, elliptic flow $v_2$, and triangular flow $v_3$.
Panel (a) indicates slight variations in central density evolution between the average and sph.-sph. scenarios in both ultracentral and semi-central collisions. 
However, these variations are insufficient to significantly impact the production of pion mesons, $v_2$, and $v_3$.

\begin{figure}[tb]
\includegraphics[width=8 cm]{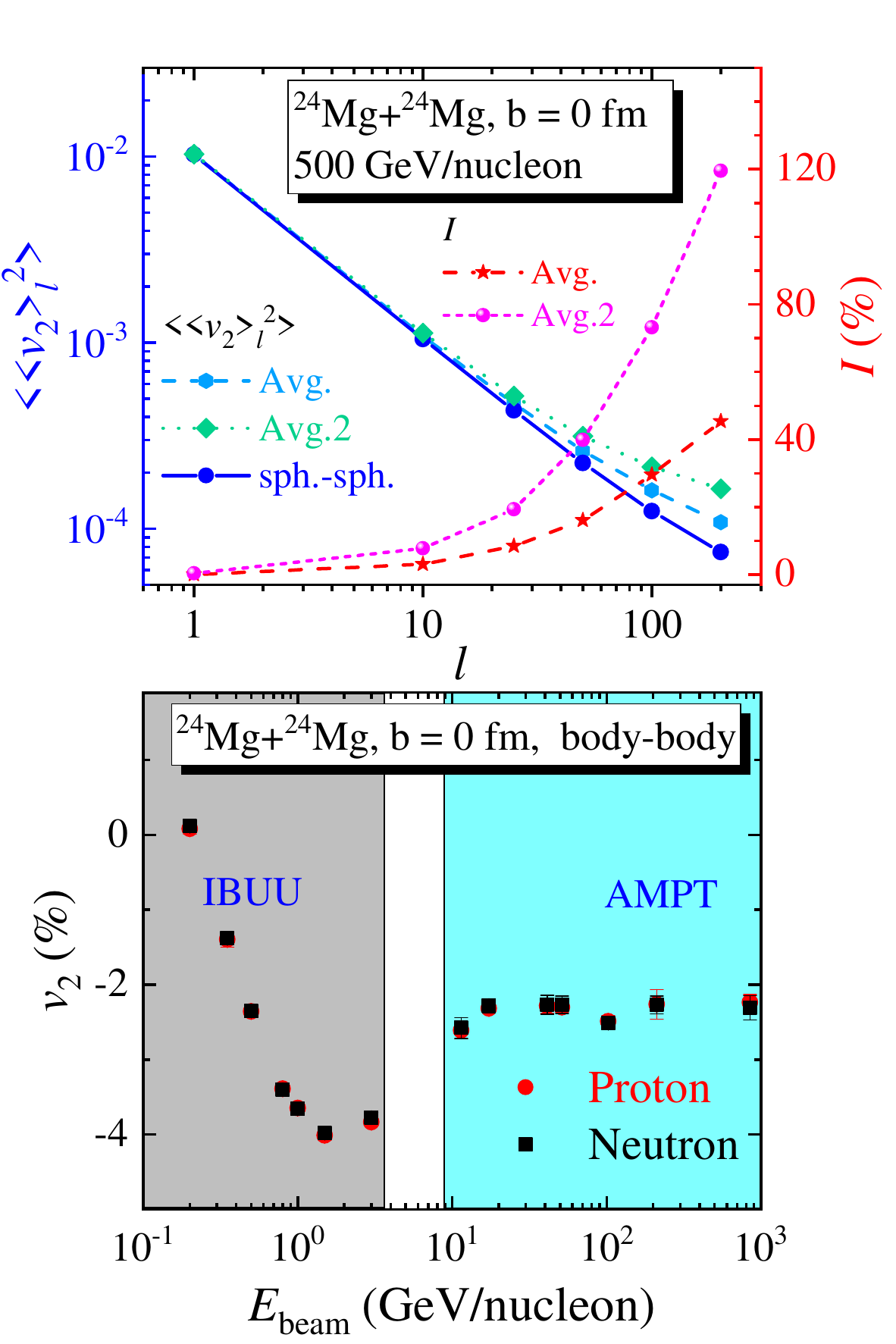}
\caption{\label{fig9} Upper panel: The mean square elliptic flows resulting from the combination of $l$ events with the same orientation (curves in the blue-green color series) and the strength of the deformation effect $I$ (curves in the red color series).
See text for details.
Lower panel: The elliptic flows produced in the body-body collision as a function of beam energy.
  }
\end{figure}

In order to investigate deformation-sensitive observables in average scenarios, we shift our focus to $\langle v_2^2 \rangle$ based on the experience gained from studying ultra-relativistic collisions. 
Here, $\langle v_2^2 \rangle$ represents the mean square elliptic flow averaged over all events, while $v_2$ denotes the elliptic flow of a single event.
In this case, it is worth noting that the number of hadrons produced in intermediate-energy ( $\leq 1 \,\text{GeV}$) reactions is often only approximately $1/20$ of that in ultra-relativistic (e.g. $\sqrt{s_{NN}} = 200 \,\text{GeV}$) reactions.
This inevitably leads to substantial fluctuations, making it challenging to capture deformation information from the results. 
To mitigate these fluctuations, the elliptic flows of $l$ events with the same orientation are combined and discussed together, described as $\langle \langle v_2 \rangle_l^2 \rangle$.
The inner and outer brackets respectively represent the averaging over $l$ events with the same orientation and averaging over all orientations.

The $\langle \langle v_2 \rangle_l^2 \rangle$ is shown (the blue-green color series) in the upper panel of Fig.~\ref{fig9} for the sph.-sph., Avg., and Avg.2 scenarios. 
The response relation now is expressed in a more general form as
\begin{equation}
    \langle \langle v_2 \rangle_l^2 \rangle = a_l + f_l(\beta),
\end{equation}
where the fluctuation $a_l$ corresponds to the blue curve and the deformation effect $f_l(\beta)$ is manifested as the difference between the green (or light blue) curve and the blue curve.
Aiming at fluctuations, extensive researches were conducted in Refs.~\cite{Giacalone2019Phys.Rev.C99.024910, Jia2022Phys.Rev.C105.014906, Zhang2022Phys.Rev.Lett.128.022301,Dimri2023Eur.Phys.J.A59.}.
One of the most direct conclusions is that the fluctuations are inversely proportional to $l$. 
As $l$ increases, the final-state hadrons under the same orientation naturally increase (with each event generating approximately 50 hadrons within the current reaction conditions), leading to decreased fluctuations. 

As fluctuations are gradually eliminated, the deformation effect becomes evident.
The strength of the deformation effect $I$ is further defined as
\begin{equation}
    I = \frac{f_l(\beta)}{a_l},
\end{equation}
which corresponds to the curves in the red color series of the panel.
When considering $ \langle v_2 ^2 \rangle$ ($l=1$), the variations caused by deformation are completely overshadowed by the fluctuation.
In the Avg. case, the strength of the deformation effect reaches a distinguishable 10\% only when $l = 25$. 
This implies that a reaction system needs to produce over 1,200 particles in a single collision to have the capability of event-by-event deformation identification.
In the intermediate-energy range, this is practically impossible to achieve.
Avg.2 brings optimism to this situation, as in the process of shielding the projectile, the orientation of the target tends to favor projecting a larger area onto the $x-y$ plane, which enhances the average deformation effect.
As such, the deformation allows being detected with $l = 10$, corresponding to 500 final-state particles. 
We can conclude that the magnitude of deformation can be probed experimentally in transport systems with projectile and target mass numbers greater than around 170.

In ultra-relativistic collisions, the deformation term can be represented as 
\begin{equation}
    f_1(\beta) = k^2 \langle\epsilon_2^2(\beta)\rangle .
\end{equation}
The $\langle\epsilon_2^2(\beta)\rangle = 0.036$ is calculated at the current density using Eq.~(\ref{eq_3}), which is nearly half of the value obtained under the Woods-Saxon density with $\langle\epsilon_2^2\rangle = 0.239 \beta_2^2 = 0.067$ in Ref.~\cite{Jia2022Phys.Rev.C105.014905}.
In this approach, we obtain $f_1(\beta) = 1.64 \times 10 ^{-4}$, where $k = 0.0657$ is taken from previous calculations using body-body collisions at 500 MeV/nucleon.
On the other hand, the $f_1(\beta) = \langle v_2^2  \rangle -  a_1 = (1.196 - 1.02) \times 10 ^{-3} = 1.76 \times 10 ^{-4}$  is further calculated at energy $\sqrt{s_{NN}} = 200\,\text{GeV}$ by using the AMPT model v2.26t5 with string-melting mode and partonic cross section of 3.0 mb \cite{Bzdak2014Phys.Rev.Lett.113.252301, Ma2014Phys.Lett.B739.209213}.
A coincidence is that the elliptic flow magnitude of body-body collisions is very similar at the two energies (see the lower panel of Fig.~\ref{fig9}), by which the similar $k$ is generated.
The close $f_1(\beta)$ values confirm the linear response relation in AMPT.
As comparisons, the corresponding results from the IBUU model are presented in Table~\ref{tab1}.

\begin{table}[h]
\centering
\caption{\label{tab1} The deformation effect $f_l(\beta)$ ($\times 10^{-5}$) in non-polarized  $^{24}\text{Mg} + ^{24}\text{Mg}$  ultracentral collisions calculated by IBUU ($E_\text{beam}$ = 0.5 GeV/nucleon) and AMPT ($\sqrt{s_{NN}} = 200\,\text{GeV}$) models.}
\renewcommand{\arraystretch}{1.2}
\begin{tabular}{c|ccccccc}
\hline\hline
$l$     & ~~~1~~~           & ~~10~~     & ~~25~~      & ~~50~~      & ~100~     & ~200~     & 1(AMPT) \\ \hline
Avg.   & 0 & 3.20 & 3.67& 3.62 & 3.67 & 3.39 & 17.6 \\ \hline
Avg.2 & 4.00       & 8.16 & 8.38 & 8.97 & 9.09 & 8.93 & ---     \\ \hline\hline
\end{tabular}
\end{table}

The deformation effect remains consistent within a narrow interval when $l\geq 10$, which is independent of $l$ and solely determined by the deformation. 
Moreover, it can be seen both Avg. and Avg.2 exhibit considerably smaller values compared to the calculations based on AMPT and $k^2 \langle\epsilon_2^2(\beta)\rangle$. 
This disparity could be attributed to the greater influence of spectators in intermediate-energy reactions.
The deformation effect is partially eliminated during the reaction process.

Through the current analysis, we can basically comprehend the geometric effects of deformed nuclei in intermediate-energy heavy-ion collisions.
As an extension, the conclusions from the prolate nucleus $^{24}\text{Mg}$ are also qualitatively applicable to some other potential prolate nuclei, such as $^{20}\text{Ne} $, etc.
When exploring other properties of nuclear systems, such as symmetry energy, neutron skin, bubble structure, and high-momentum tail, this study can also serve as a reference for eliminating the interference of deformation effects.
It should be noted that although the density difference caused by the nuclear forces in this study is small enough not to significantly affect the observables, it remains a worthwhile direction for future research to introduce relativistic mean-field interactions in the IBUU model to maintain consistency.

On another aspect, the high sensitivity of collective flows to deformation provides a promising prospect for studying deformation based on neural network methods, by which density-related studies have already achieved fundamental progress \cite{Yang2023Phys.Lett.B840.137870, Yang2021Phys.Lett.B823.136650}.
Constructing deep-neural-network mappings from collective flows $(v_2^2, v_3^2, ...)$ to densities $(\rho_n, \rho_p)$ aimed at certain centralities will allow the straightforward generation of nuclear density profiles through experimental data.
This will further deepen our knowledge of nuclear structures.

Based on this study, further systematic research on EoS can be conducted. 
Comparing the theoretical and experimental $v_2$ signals at central body-body collisions between $^{24}\text{Mg}$ to deduce EoS has explicit advantages: 
1.~shadowing effects, which is the squeeze-out effect by the spectators at mid-rapidity region \cite{Petersen2006Phys.Rev.C74.064908}, can be avoided to some extent as all the nucleons are participants;
2.~the intermediate-energy $v_2$ signal can be detected experimentally in a more economical way at HIAF/IMP, J-PARC, FAIR, NICA, HIMAC, etc.
This implies that we also need to understand thoroughly the fundamental mechanisms behind squeeze-out effects and in-plane flow during deformed nuclei reactions.
To summarize, a more accurate extraction of the EoS dependence of elliptic flow is anticipated in ultracentral deformed nucleus-nucleus collisions.

\section{summary \label{sec:6}}

Heavy-ion collision simulations of deformed nuclei based on the IBUU model are conducted, where the highly deformed prolate nucleus $^{24}$Mg is comprehensively studied. 
The spatial density distributions obtained from RMF calculations are used to initialize the tip-tip and body-body collisions, compared with the sph.-sph. collisions initialized via the constraint calculation on $\beta_2 = 0$ point.

Regarding these three collision scenarios, we first explore the role played by centrality. 
From the central baryon density, we notice that in tip-tip and sph.-sph. collisions, the reaction pressure and reaction time decrease significantly as the centrality decreases. 
This is in contrast to body-body collisions, which are less sensitive to centrality within the domain $b \in [0,6] \, \text{fm}$.
As a result, the discrimination in reaction pressure is directly reflected in the multiplicities of mesons and nucleons, where the meson yield of body-body collisions is roughly twice that of sph.-sph. collisions and three times that of tip-tip collisions in peripheral collisions.
During the study of elliptic flow, it is noted that an increasing impact parameter can enhance the flows for tip-tip and sph.-sph. collisions, while the body-body case still maintains a relatively small variation.
We attribute the anomaly in the body-body case to the dilute participants in the periphery of the systems.

The study on the collision angle is further carried out, exploring the corresponding particle yield and elliptic flow variations by adjusting the orientation of the projectile in $15^\circ$ intervals. 
Obviously, the initial eccentricity of the overlap $\epsilon_{2}$ still dominates the variation of the elliptic flow $v_2$, maintaining a positive correlation.
Subsequently, the special case of the tip-body is discussed further, where triangular flow expressly emerges at high $p_t$ on non-central collisions.
The reliability of $v_3$ as a probe for nuclear deformation and the reaction centrality of quadrupole-deformed nuclei is thus recognized.

In the final analysis, the scenarios of non-polarized collisions are implemented. 
The reaction results reveal a relatively subdued mean squared elliptic flow compared to the predictions of the linear response relation $k^2 \langle\epsilon_2^2(\beta)\rangle$ and the outcomes in AMPT at relativistic energies.
By filtering out events with spectators from the projectiles, there is a noticeable increase in the mean squared elliptic flow. 
However, the event-by-event analysis of deformation information remains unattainable. 
It is speculated that deformation effects in intermediate-energy reactions only manifest when both the mass numbers of the projectile and target exceed 170 at least.

This work will also contribute to the studies about intermediate-energy heavy-ion collisions between other deformed nuclei, and facilitate further discussions on properties such as symmetry energy and high-momentum tails with deformed nuclei.
On the other hand, further studies on higher-order deformation components are also expected.

\section{Acknowledgements}

We acknowledge helpful discussions with Profs.~Haozhao Liang, Gao-Chan Yong, Jian Xiang, and Wei Sun. 
This work is supported by the National Natural Science Foundation of China under Grants Nos. 12005175, 11875225.
the Fundamental Research Funds for the Central Universities under Grant No. SWU119076, the JSPS Grant-in-Aid for Early-Career Scientists under Grant
No. 18K13549, the JSPS Grant-in-Aid for Scientific Research (S) under Grant No. 20H05648. This work was also partially supported by the RIKEN Pioneering
Project: Evolution of Matter in the Universe and the INT Program INT-23-1a and Institute for Nuclear Theory.

\bibliographystyle{apsrev4-1}
\bibliography{Ref}

\end{document}